\documentclass[twocolumn,twoside,slac_two]{revtex4}
\usepackage{graphicx}
\usepackage{fancyhdr}
\begin{document}

\title{Measurement of $B \to D^{*-} \tau^+ \nu_\tau$ and $B \to h^{(*)} \nu\overline{\nu}$ Decays at Belle}

%

\author{K.-F. Chen}
\affiliation{Department of Physics, National Taiwan University, Taipei}

\begin{abstract}
We report an observation of the decay $B^0 \to D^{*-} \tau^+ \nu_{\tau}$ and 
a search for the rare decays $B \to h^{(*)} \nu \overline{\nu}$, where $h^{(*)}$ stands for a light meson.
A data sample of 535 million $B\overline{B}$ pairs 
collected with the Belle detector at the KEKB $e^+e^-$ collider is used. 
We find a signal with a significance of 5.2 standard deviations 
on $B^0 \to D^{*-} \tau^+ \nu_{\tau}$ and measure the branching fraction
to be $2.02^{+0.40}_{-0.37}({\rm stat.}) \pm 0.37({\rm syst.})$\%.
No significant signal is observed for $B \to h^{(*)} \nu \overline{\nu}$ decays and 
we set upper limits on the branching fractions at 90\% confidence level.
The limits on $B^0 \to K^{*0}\nu\overline{\nu}$ and $B^+ \to K^+\nu\overline{\nu}$ decays
are more stringent than the previous constraints, while the first searches for
$B^0 \to K^0 \nu\overline{\nu}$, $\pi^0\nu\overline{\nu}$, $\rho^0\nu\overline{\nu}$, $\phi\nu\overline{\nu}$ and
$B^+ \to K^{*+}\nu\overline{\nu}$, $\rho^+\nu\overline{\nu}$ are presented.
\end{abstract}

\maketitle

\thispagestyle{fancy}


\section{Introduction}

The decay $B^0 \to D^{*-} \tau^+ \nu_{\tau}$ is dominated 
by the $b \to c$ transition and can provide the 
important information associated with the charge Higgs in the Standard Model (SM).
The $\tau$ lepton in the final state provide additional 
observables sensitive to the physics beyond SM, as well as the $\tau$ polarization,
which cannot be accessed in other semileptonic decays.
However, the neutrinos in the final states and the low 
efficiencies from $\tau$ reconstruction make the search
to be very challenging.
The SM predict a $B \to \overline{D}^{*} \tau^+ \nu_{\tau}$ branching fraction of 
1.4\%~\cite{ref:dtaunu_bf}, while there are several experimental results provided by the LEP experiments;
the averaged $b \to \tau \nu_{\tau} X$ semi-inclusive branching fraction is
$2.48\pm0.26$\%~\cite{ref:PDG06}.

The flavor-changing neutral-current process $B \to h^{(*)} \nu \overline{\nu}$ is
sensitive to physics beyond the SM. 
The SM branching fractions are estimated to be 
$1.3 \times 10^{-5}$ and $4 \times 10^{-6}$ for $B \to K^* \nu \overline{\nu}$
and $B \to K \nu \overline{\nu}$ decays~\cite{ref:buchalla}, respectively, 
and are expected to be much lower for other modes.
Theoretical calculation of the decay amplitudes for these decays
is particularly reliable, because of the absence of long-distance
interactions that affect charged-lepton channels $B \to h^{(*)} l^+l^-$.
New physics such as SUSY particles or a possible fourth generation could
potentially contribute to the penguin loop or box diagram 
and enhance the amplitudes~\cite{ref:buchalla}.
Reference~\cite{ref:darkmatter} also discusses the possibility of
discovering light dark matter in $b\to s$ transitions with large
missing momentum.
Due to the challenge of cleanly detecting rare modes with two final-state
neutrinos, only a few studies of $h^{(*)}\nu \overline{\nu}$ have been
carried out to date~\cite{Adam:1996ts,Browder:2000qr,Aubert:2004ws}.

In this report, we present 
the first observation of the decay $B^0 \to D^{*-} \tau^+ \nu_{\tau}$ 
and the search for the decays $B \to h^{(*)} \nu \overline{\nu}$ 
($h^{(*)}$ stands for $K^+$, $K_S^0$, $K^{*0}$, $K^{*+}$, $\pi^+$, $\pi^0$, $\rho^0$, $\rho^+$, and $\phi$) 
using a 492 fb$^{-1}$ data sample recorded at the $\Upsilon(4S)$ resonance, 
corresponding to $535\times 10^6$ $B$-meson pairs. Throughout this report, 
the inclusion of charge conjugate decays is implied unless otherwise stated.
The Belle detector is a large-solid-angle magnetic spectrometer located at
the KEKB collider~\cite{ref:KEKB}, and is described in detail
elsewhere~\cite{ref:belle_detector}.

\section{$B^0 \to D^{*-} \tau^+ \nu_{\tau}$}

We select charged tracks that
are associated with the interaction point (IP). 
The electrons candidates are selected
using the information from particle identification
systems. 
The four momenta of electron candidates
are corrected for bremsstrahlung radiation
by adding photons within a 50 mrad cone along the track direction.
The $\pi^0$ candidates are reconstructed from
pairs of photon with the invariant mass in the range 118 MeV/$c^2$
and 150 MeV/$c^2$. Minimum energies of 60--120 MeV
are required for the photon canidates from $\pi^0$ decays, according to 
different polar angles. 
Photons that are not included in $pi^0$ reconstruction
and exceed a polar-angle dependent energy threshold
(100--200 MeV) are included in the tag-side $B$-meson ($B_{\rm tag}$) reconstruction.

Reconstruction of the $B_{\rm tag}$ strongly suppresses
the combinatorial and continuum backgrounds
and provides kinematical constraints on the signal meson
($B_{\rm sig}$). We take the advantage of the clean
signature, supported by the $D^{*}$ meson at the signal side.
The $B_{\rm tag}$ meson is reconstructed using  
all the particles that remain after selecting candidates
for $B_{\rm sig}$ decay daughters. 
The $D^{*}$ mesons are reconstructed through the following decay
chain: $D^{*}+ \to D^0 \pi^+$, $D^0 \to K^- \pi^+$ and $K^-\pi^+\pi^0$.
The $\tau$ leptons are reconstructed in $\tau \to e^+ \nu_e \overline{\nu}_\tau$
and $\pi^+\overline{\nu}_\tau$ decays, while the $\tau \to \mu^+ \nu_e \overline{\nu}_\tau$
mode is excluded due to the inefficient muon identification
in the relevant momentum range.
For $\tau^+ \to \pi^+\overline{\nu}_\tau$ decays, only $D^0 \to K^- \pi^+$
mode is used in order to avoid the higher combinatorial background.

Once a $D^{*+}$ candidate is reconstructed and a charged track expected from $\tau^+$ is selected, the
remaining particles measured by the detector are used to reconstruct the $B_{\rm tag}$.
Two kinematical variables, $M_{\rm tag} = \sqrt{E_{\rm beam}^2 - p_{\rm tag}^2}$ and 
$\Delta E_{\rm tag} = E_{\rm tag} - E_{\rm beam}$, are used to identify the 
$B_{\rm tag}$ candidates, where $E_{\rm beam}$ is the beam energy. 
The momentum ($p_{\rm tag}$) and energy ($E_{\rm tag}$) of the  $B_{\rm tag}$
meson is calculated by a summation over
all particles that are not assigned to $B_{\rm sig}$.
The signal candidates are required to satisfiy 
$M_{\rm tag} > 5.2$ GeV/$c^2$ and $|\Delta E_{\rm tag}|<0.6$ GeV at least.
To improve the purity of the selected $B_{\rm tag}$ candidates,
several additional requirements are imposed, such as 
zero total event charge, no additional leptons in the event, zero
bayron number. The residual energy in the ECL should be smaller than
0.35 GeV and number of neutral particles ($\pi^0$ and $\gamma$) included
for the tag side should be less then 5.
The $B_{\rm tag}$ reconstruction algorithm is varified using 
the control sample, $B_{\rm sig} \to D^{*-}\pi^+$, and is found to be 
consistent with Monte Carlo (MC) simulations.

The dominated background source is from the semileptonic $B \to D^{*} e \nu_{e}$ decays
for $\tau^+ \to e^+ \nu_e \overline{\nu}_\tau$ mode, and combinatorial background 
from hadonic $B$ decays for $\tau^+ \to \pi^+\overline{\nu}_\tau$ decays.
Further background suppression is achieved with the following variables:
the missing energy $E_{\rm mis} = E_{\rm beam} - E_{D^*} - E_{e,\pi}$, visible
energy of the event, the square of missing mass $M_{\rm mis}^2 = E_{\rm mis}^2 - (p_{\rm sig} - p_{D^*} -  p_{e,\pi})^2$,
and the effective mass of $\tau\nu_{\tau}$ system $M_W^2 = (E_{\rm beam} - E_{D^*})^2 - (p_{\rm sig}- p_{D^*})^2$. 
The most effective variable $X_{\rm mis}$ is defined by $(E_{\rm mis} - |p_{D^*}+p_{e,\pi}|)/\sqrt{E_{\rm beam}^2 - m_{B^0}^2}$,
which is closely related to the missing mass but does not depend on $B_{\rm tag}$ reconstruction.

\begin{figure}[h]
\centering
\includegraphics[width=60mm]{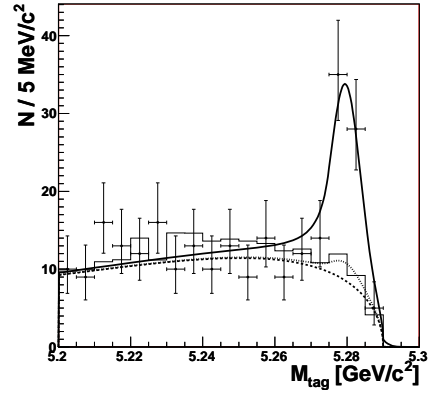}
\caption{$M_{\rm tag}$ and $\Delta E_{\rm tag}$ distributions for $B \to D^{*-} \tau^+ \nu_\tau$ candidates from data. 
The solid curve shows the fit results, and the dot-dashed curves indicate the background component.
The open-histograms shows the background distributions from MC simulations.} \label{fig:dtaunu}
\end{figure}

We extract the signal yields by maximum likelihood fits to the $M_{\rm tag}$ distributions.
The likelihood function is given by
\begin{equation}
\mathcal{L} = e^{-(N_s+N_p+N_b)} \prod_{i=1}^N [(N_s+N_p)P_s(M_{\rm tag}^i)+N_bP_b(M_{\rm tag}^i)]~,
\end{equation}
where $P_s$ ($P_b$) is the probability density function (PDF) for signal and combinatorial background events,
and $N_s$, $N_p$, and $N_b$ denote the yields for signal, pecking background, and combinatorial background,
respectively. The signal distribution is described using
a Crystal Ball lineshape function~\cite{ref:cbline}, and the background part is parameterized using
the ARGUS-function~\cite{ref:argus}. The number of $N_s$ and $N_b$ are float parameters in the fit, while 
the $N_p$ is fixed to the value obtained from MC simulation, and fixed to zero for 
 $\tau^+ \to \pi^+\overline{\nu}_\tau$ decays. 
The fit results are included in Table~\ref{tab:dtaunu}, and the distributions of $M_{\rm tag}$ and 
$\Delta E_{\rm tag}$ from data with fit results superimposed are shown in Figure~\ref{fig:dtaunu}.
The combined branching fraction is $2.02^{+0.40}_{-0.37}$\%, and is obtained using a fit with a constraint to a common value.

\begin{table}[h]
\begin{center}
\caption{Summary of signal yield ($N_s$), reconstruction efficiencies ($\epsilon$), 
branching fraction ($\mathcal{B}$), and statistical significance ($\Sigma$) for 
$B \to D^{*-} \tau^+ \nu_\tau$ decays.}
\begin{tabular}{|l|c|c|c|c|}
\hline \textbf{subchannel} & $\mathbf{N_s}$ & $\mathbf{\epsilon(10^{-4})}$ & $\mathcal{B}\mathbf{(\%)}$ & $\mathbf{\Sigma}$ \\
\hline 
$D^0 \to K^- \pi^+$, &
$19.5^{+5.8}_{-5.0}$ & 3.25 & $2.44^{+0.74}_{-0.65}$ & 5.0 \\
$\tau^+ \to e^+ \nu_e \overline{\nu}_\tau$  & & & & \\
\hline 
$D^0 \to K^- \pi^+ \pi^0$, &
$11.9^{+6.0}_{-5.2}$ & 0.78 & $1.69^{+0.84}_{-0.74}$ & 2.6 \\
$\tau^+ \to e^+ \nu_e \overline{\nu}_\tau$  & & & & \\
\hline 
$D^0 \to K^- \pi^+$, &
$29.9^{+10.0}_{-9.1}$ & 1.07 & $2.02^{+0.68}_{-0.61}$ & 3.8 \\
$\tau^+ \to \pi^+ \overline{\nu}_\tau$  & & & & \\
\hline 
Combined &
$60^{+12}_{-11}$ & 1.17 & $2.02^{+0.40}_{-0.37}$ & 6.7 \\
\hline
\end{tabular}
\label{tab:dtaunu}
\end{center}
\end{table}

The systematic uncertainties include the number of $B$-meson pairs (1.3\%), 
signal shape (2.8\%), parameterization of the combinatorial background (5.7\%) 
which is estimated by varying the ARGUS-shape parameters.
An 8.2\% uncertainty is included for the peaking background,
which is dominated by MC statistics.
The uncertainty in $B_{\rm tag}$ reconstruction (10.9\%) is evaluated from the control sample.
Efficiency uncertainties in the
tracking, neutral reconstruction and
particle identification are in the range of 7.9--10.7\%, 
according to different decay channels.
The uncertainties due to the partial sub ratios are taken from PDG~\cite{ref:PDG06}.
The combined uncertainty is 18.5\%, and the statistical significance 
signal is reduced to 5.2$\sigma$ including the systematic uncertainties.

In conclusion,we observe $60^{+12}_{-11}$ events for the decay $B^0 \to D^{*-} \tau^+ \nu_{\tau}$
based on a data sample of $535\times 10^6$ $B\overline{B}$ pairs.
This is the first observation of an exclusive $B$ decays with
$b \to c \tau \overline{\nu}_\tau$ transition. The measured branching fraction
$2.02^{+0.40}_{-0.37}\pm0.37$\% is consistent with the prediction in SM.

\section{$B \to h^{(*)} \nu\overline{\nu}$}

The decays $B \to h^{(*)} \nu\overline{\nu}$ are reconstructed in a different way.
Candidate $e^+e^- \to \Upsilon(4S) \to B\overline{B}$ 
events are characterized by a fully-reconstructed $B_{\rm tag}$.
The $B_{\mathrm{tag}}$ candidates are reconstructed in one of the following modes: $B^0 \to
D^{(*)-} \pi^+$, $D^{(*)-}\rho^+$, $D^{(*)-}a_1^+$, $D^{(*)-}D_s^{(*)+}$, 
$B^+ \to \overline{D}{}^{(*)0} \pi^+$, $\overline{D}{}^{(*)0} \rho^+$,
$\overline{D}{}^{(*)0} a_1^+$, and $\overline{D}{}^{(*)0} D_s^{(*)+}$.
The $D^-$ mesons are reconstructed as $D^- \to K^0_S\pi^-$, $K_S^0\pi^-\pi^0$, 
$K_S^0\pi^-\pi^+\pi^-$, $K^+\pi^-\pi^-$, and $K^+\pi^-\pi^-\pi^0$. 
The following decay channels are included for $\overline{D}{}^0$ mesons: 
$\overline{D}{}^0 \to K^{+}\pi^{-}$, $K^+\pi^-\pi^0$, $K^+\pi^-\pi^+\pi^-$, 
$K_S^0\pi^0$, $K_S^0\pi^-\pi^+$, $K_S^0\pi^-\pi^+\pi^0$ and $K^-K^+$. 
The $D^{*-}$ ($\overline{D}{}^{*0}$) mesons are reconstructed as $\overline{D}{}^0 \pi^-$ 
($\overline{D}{}^0 \pi^0$ and $\overline{D}{}^0 \gamma$).
Furthermore, $D_s^{*+} \to D_s^+ \gamma$, $D_s^+\to K_S^0K^+$ and $K^+K^-\pi^+$ decays
are reconstructed. 
$B_{\mathrm{tag}}$ candidates are selected using the
beam-energy constrained mass $M_{\mathrm{bc}} \equiv \sqrt{E_{\mathrm{beam}}^{2} - p_{B}^{2}}$ 
and the energy difference $\Delta E \equiv E_{B} - E_{\mathrm{beam}}$.
We require $B_{\mathrm{tag}}$ candidates satisfy
the requirements 
$M_{\mathrm{bc}}>5.27$~GeV/$c^2$ and $-80$~MeV~$<\Delta E< 60$~MeV.
We reconstruct $7.88 \times 10^5$
and $4.91 \times 10^5$ charged and
neutral $B$ mesons, respectively.

The rest of particles in the event are used to reconstruct a $B_{\rm sig} \to h^{(*)}\nu\overline{\nu}$ candidate.
Prompt charged tracks are required to associated with IP, and a minimum momentum of 0.1 GeV/$c$ in the transverse plane.
We select kaon and pion from charged tracks based on the particle identification system.
Pairs of oppositely charged tracks are used
to reconstruct $K^0_S \to \pi^+\pi^-$ decays.
For $\pi^{0} \to \gamma\gamma$, a minimum photon energy of 50 MeV
is required and the $\gamma\gamma$ invariant mass must be within $\pm$16 MeV/$c^2$ of the nominal $\pi^{0}$ mass.

The decays $B_{\mathrm{sig}} \rightarrow K^+ \nu \overline{\nu}$, $\pi^+\nu\overline{\nu}$, $K_S^0 \nu\overline{\nu}$, and 
$\pi^0 \nu\overline{\nu}$ are reconstructed from single $K^+$, $\pi^+$, $K_S^0$, and $\pi^0$ candidates, respectively.
The $B^0 \rightarrow K^{*0}\nu\overline{\nu}$ candidate is reconstructed from a charged pion and 
an oppositely charged kaon,
while $B^+ \rightarrow K^{*+}\nu\overline{\nu}$ decays are reconstructed from a $K_S^0$ candidate and a charged pion, or
a charged kaon and a $\pi^0$ candidate. 
The reconstructed mass of the $K^{*0}$ ($K^{*+}$) candidate should be within a $\pm$75 MeV$/c^{2}$ window around
the nominal $K^{*0}$ ($K^{*+}$) mass. Furthermore, pairs of charged pions with opposite charge are used to form 
$B^0 \rightarrow \rho^0 \nu\overline{\nu}$ candidates where the $\pi^+\pi^-$ invariant mass should be 
within $\pm$150 MeV/$c^2$ from the nominal $\rho^0$ mass. For $B^+ \rightarrow \rho^+\nu\overline{\nu}$, 
a charged pion and a $\pi^0$ candidate are used, and a $\pm$150 MeV/$c^2$ mass window is required.
A $\phi$ meson is formed from a $K^+K^-$ pair with a 
reconstructed mass within $\pm$10 MeV/$c^2$ from the nominal $\phi$ mass.

We reject the events with additional charged tracks or $\pi ^{0}$ candidates,
and select $B_{\mathrm{sig}}$ candidates using
the variable $E_{\mathrm{ECL}}\equiv E_{\mathrm{tot}}-E_{\mathrm{rec}}$,
where $E_{\mathrm{tot}}$ and $E_{\mathrm{rec}}$ are the total visible energy
measured by the ECL detector and the measured energy of reconstructed objects
including the $B_{\mathrm{tag}}$ and the signal side $h^{(*)}$ candidate,
respectively. 
The decays $B \to D^{*} \ell \nu$ are examined as control samples;
the observed $E_{\mathrm{ECL}}$ distributions are found to be in good agreement with MC simulations.
The signal region is defined by $E_{\mathrm{ECL}}$~$<$~0.3~GeV
while the sideband region is given by 0.45~GeV~$<E_{\mathrm{ECL}}<$~1.5~GeV.

The dominant background source is 
$B\overline{B}$ decays involving a $b \to c$ transition.
A lower bound of 1.6 GeV/$c$
on $P^*$, the momentum of the $h^{(*)}$ (except $\phi$) in the $B_{\mathrm{sig}}$ rest frame,
suppresses this background, while an upper bound of 2.5 GeV/$c$ rejects
the contributions from radiative two-body modes such as $B \to K^*\gamma$.
The cosine of the angle between the missing momentum in the
laboratory frame and the beam is required to 
lie between $-0.86$ and $0.95$.
Other background sources are found to be small. 

\begin{figure}[htpb]
\begin{center}
\unitlength=1cm
\begin{picture}(8.0,2.8)
\put( 0.0,0.0){\includegraphics[width=4.cm,height=2.8cm]{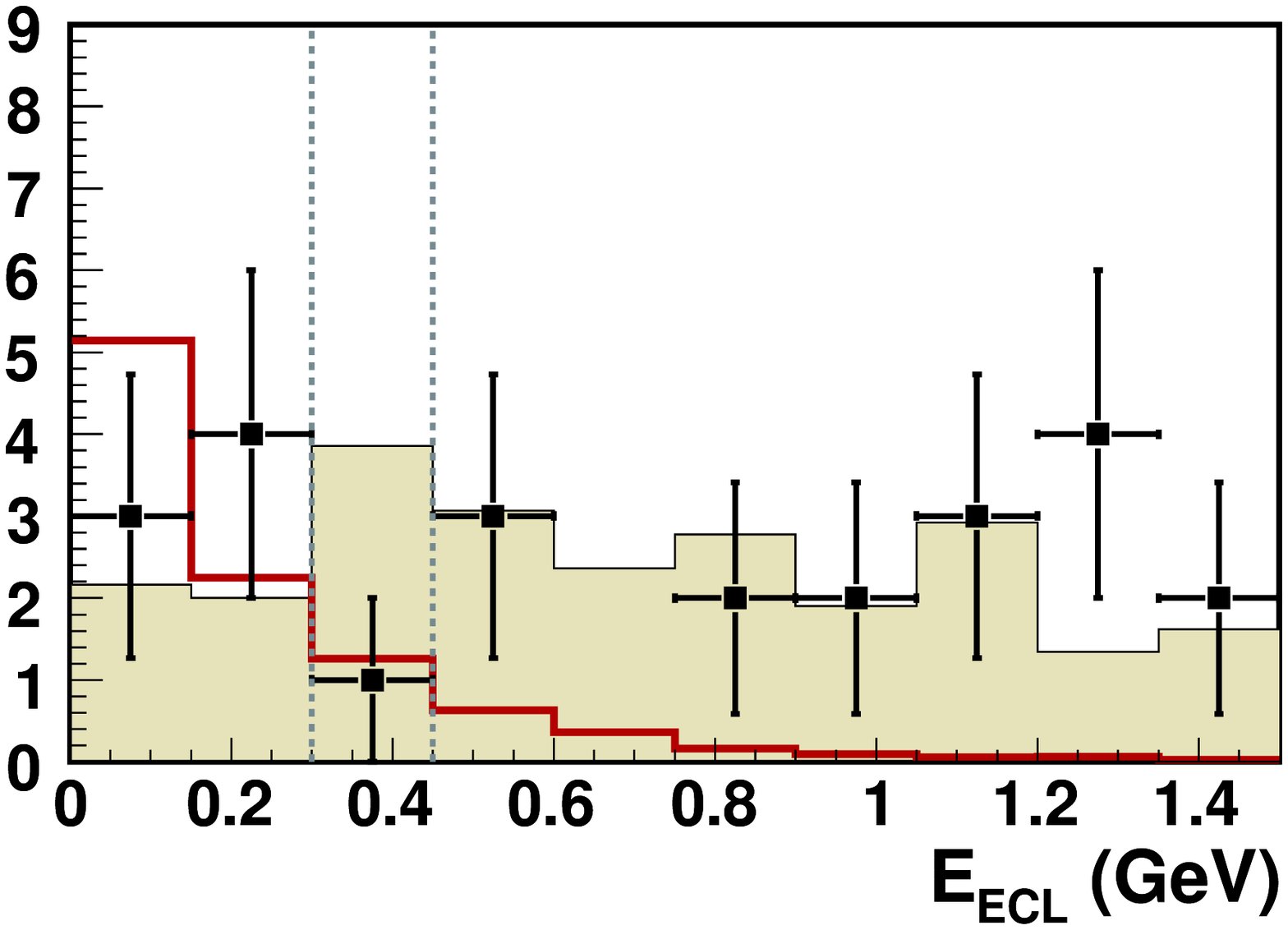}}
\put( 4.0,0.0){\includegraphics[width=4.cm,height=2.8cm]{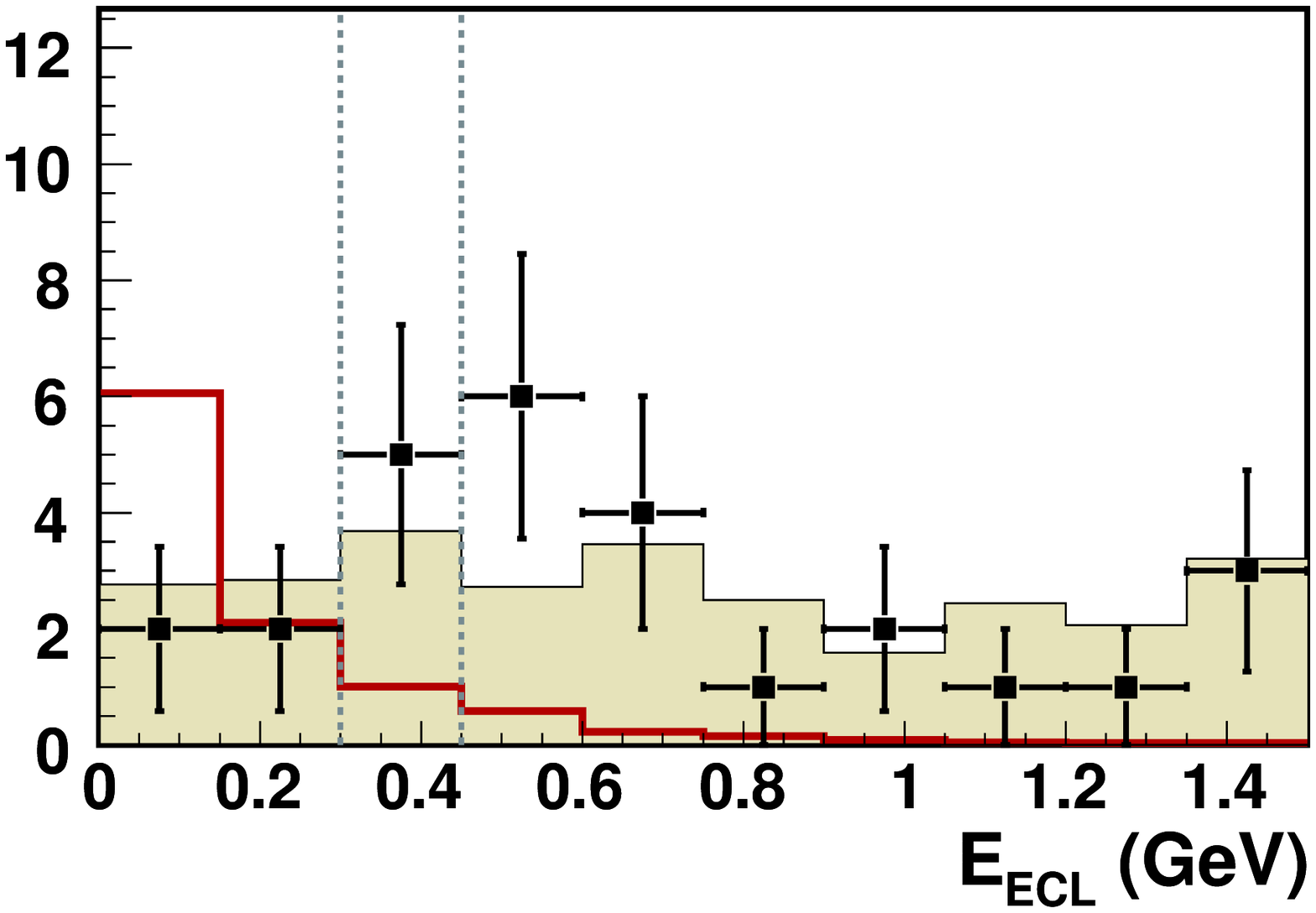}}
\put( 1.3,2.3){\footnotesize a) $B^0 \to K^{*0} \nu \overline{\nu}$}
\put( 5.3,2.3){\footnotesize b) $B^+ \to K^{*+} \nu \overline{\nu}$}
\end{picture}
\begin{picture}(8.0,2.8)
\put( 0.0,0.0){\includegraphics[width=4.cm,height=2.8cm]{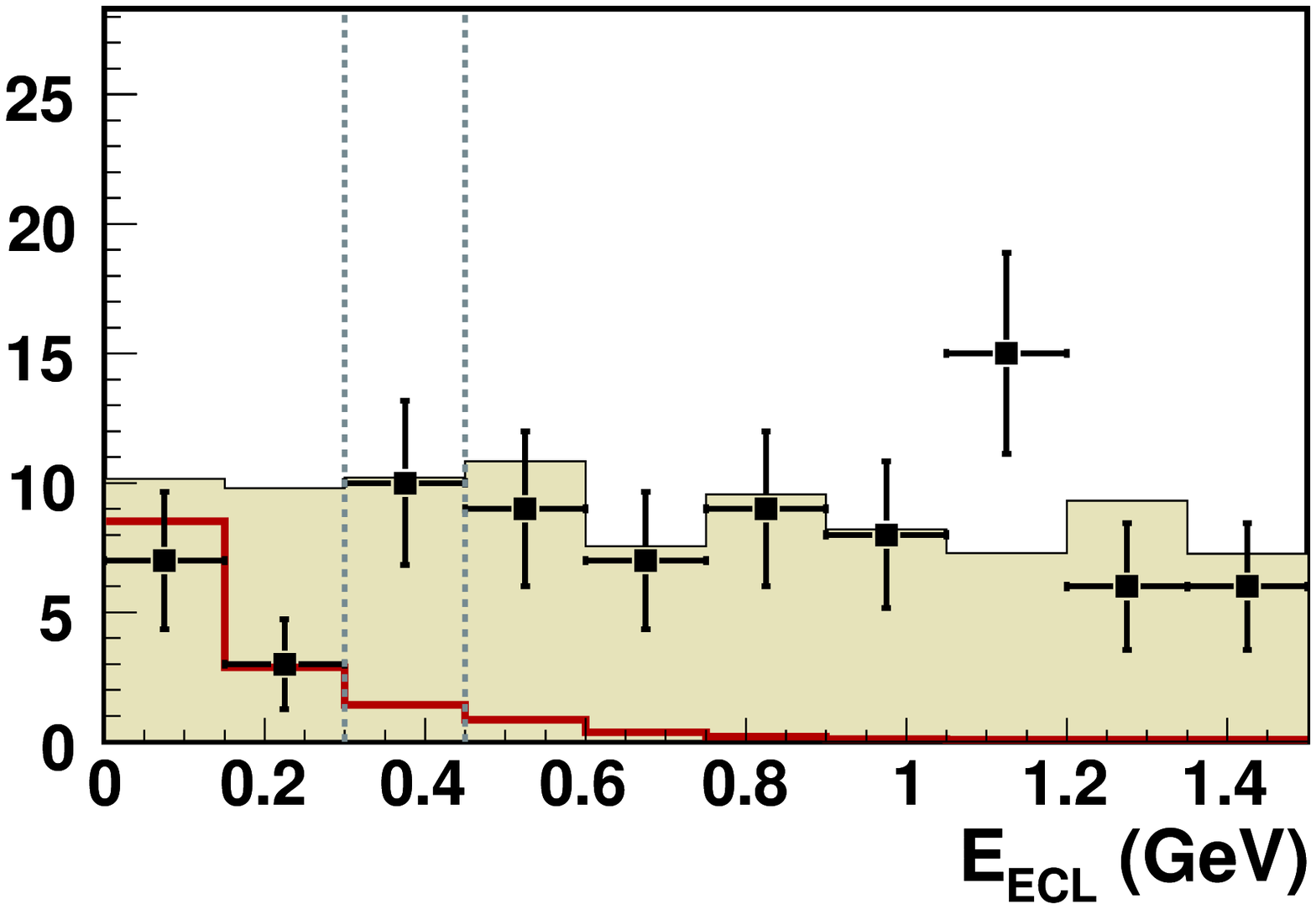}}
\put( 4.0,0.0){\includegraphics[width=4.cm,height=2.8cm]{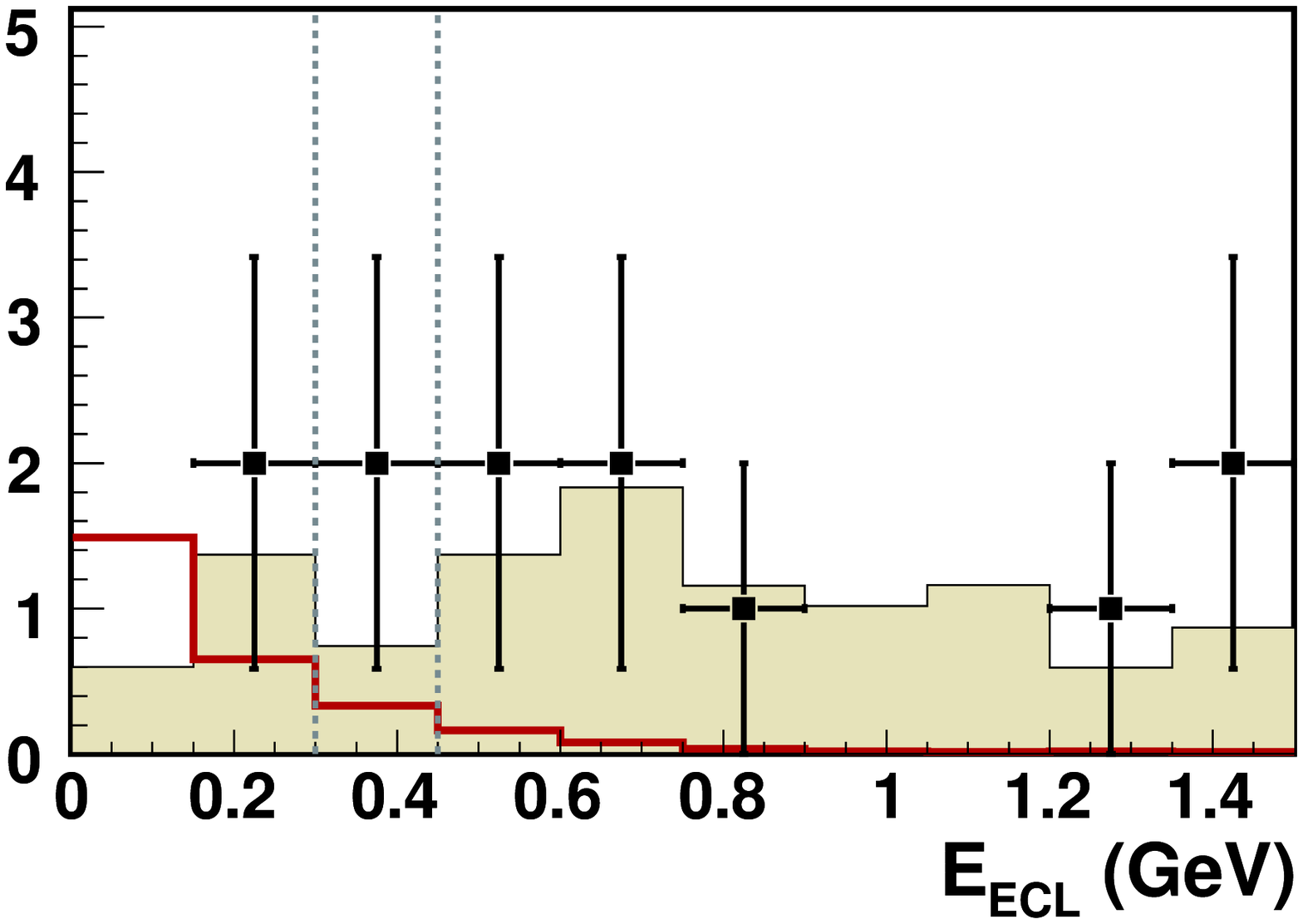}}
\put( 1.3,2.3){\footnotesize d) $B^0 \to K_S^0\nu \overline{\nu}$}
\put( 5.3,2.3){\footnotesize c) $B^+ \to K^+  \nu \overline{\nu}$}
\end{picture}
\begin{picture}(8.0,2.8)
\put( 0.0,0.0){\includegraphics[width=4.cm,height=2.8cm]{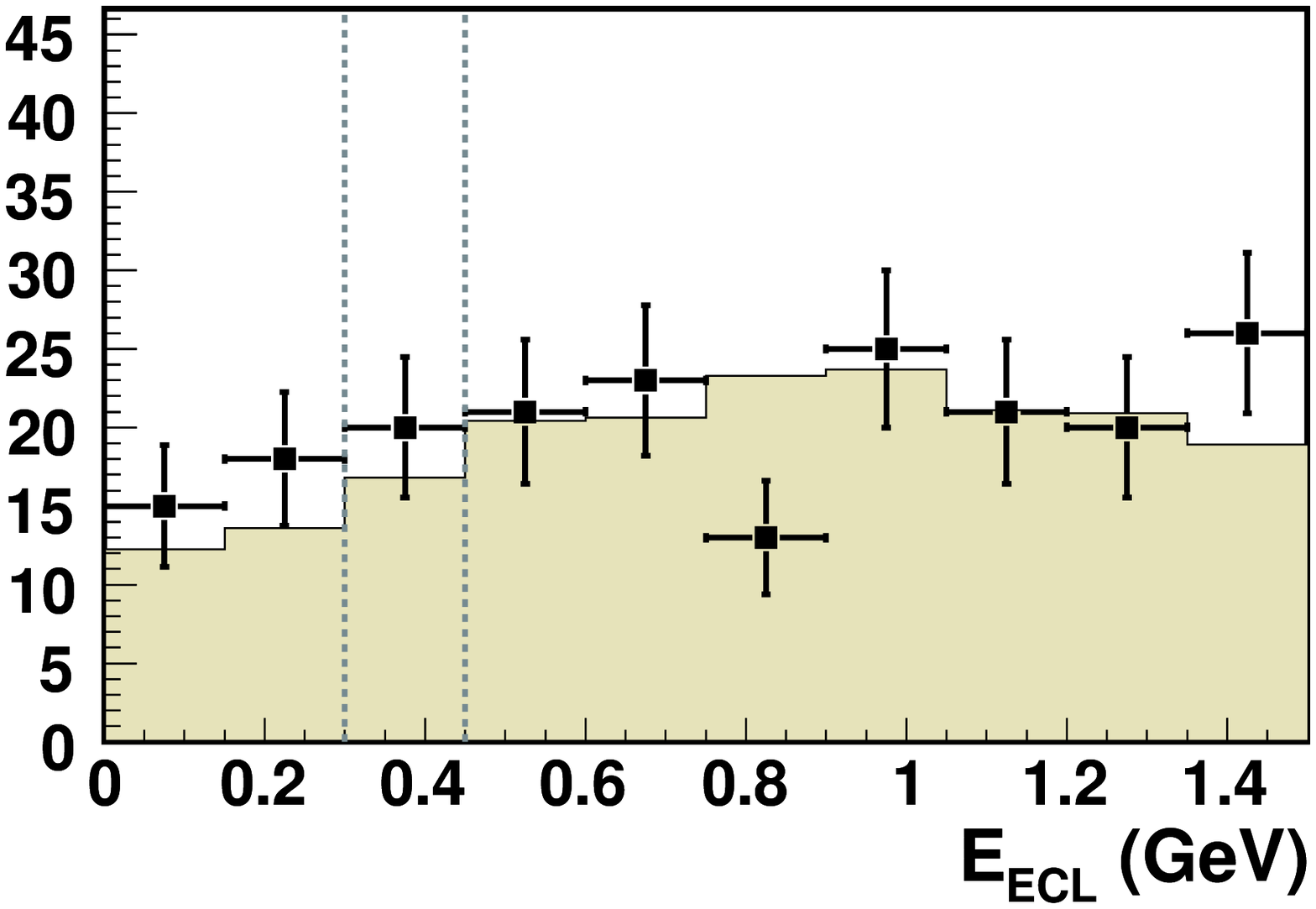}}
\put( 4.0,0.0){\includegraphics[width=4.cm,height=2.8cm]{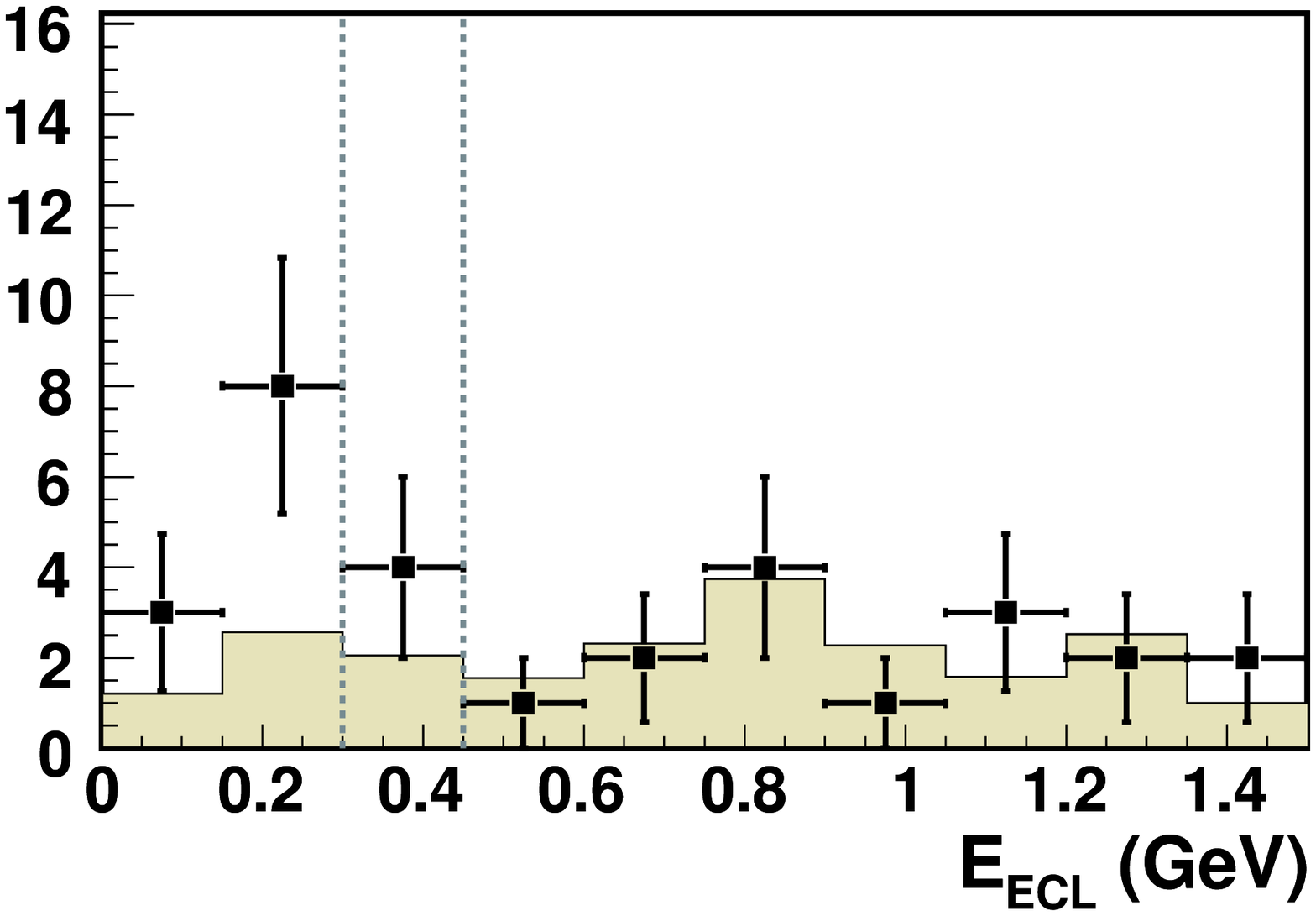}}
\put( 1.3,2.3){\footnotesize e) $B^+ \to \pi^+ \nu \overline{\nu}$}
\put( 5.3,2.3){\footnotesize f) $B^0 \to \pi^0 \nu \overline{\nu}$}
\end{picture}
\begin{picture}(8.0,2.8)
\put( 0.0,0.0){\includegraphics[width=4.cm,height=2.8cm]{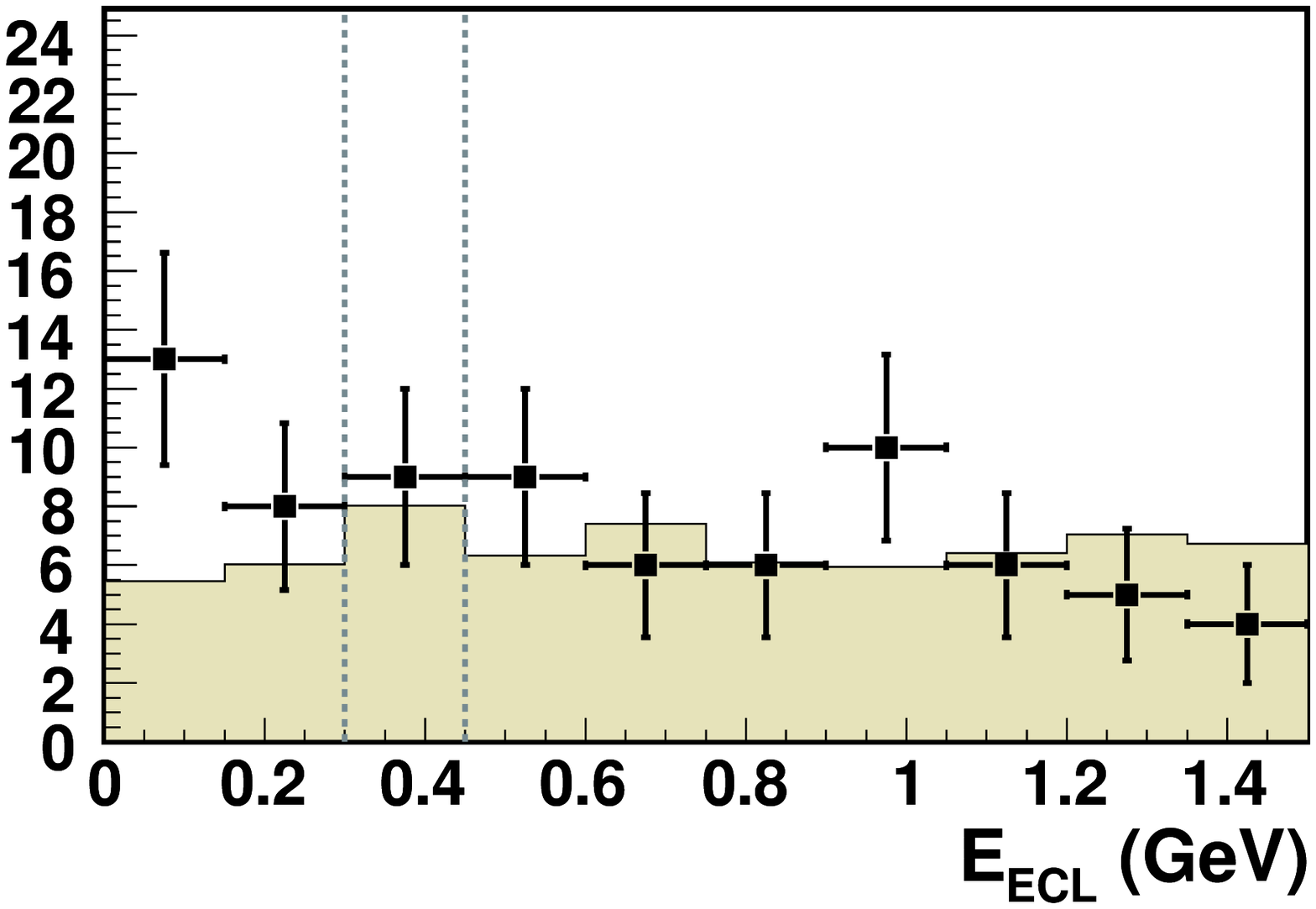}}
\put( 4.0,0.0){\includegraphics[width=4.cm,height=2.8cm]{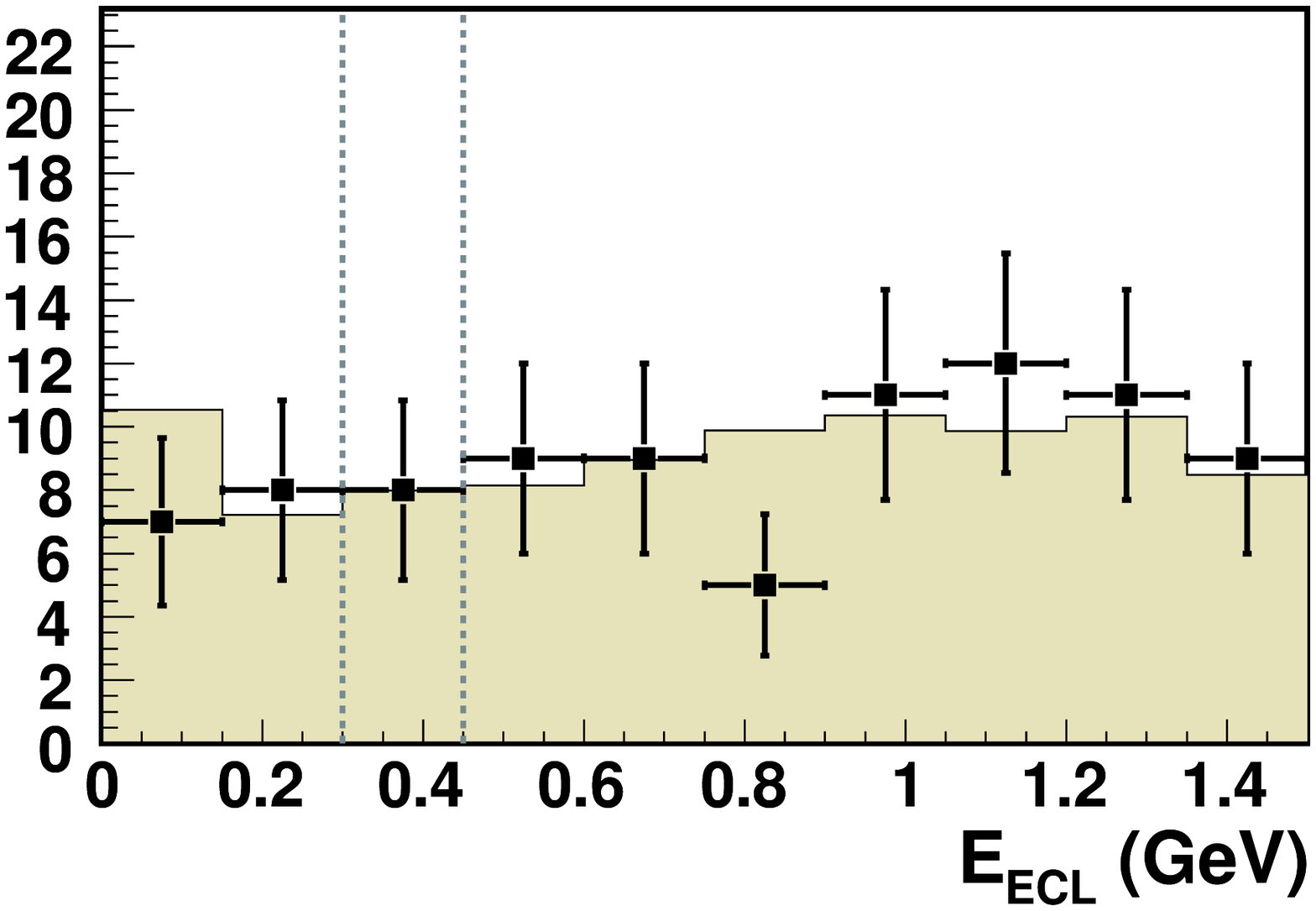}}
\put( 1.3,2.3){\footnotesize g) $B^0 \to \rho^0 \nu \overline{\nu}$}
\put( 5.3,2.3){\footnotesize h) $B^+ \to \rho^+ \nu \overline{\nu}$}
\end{picture}
\begin{picture}(8.0,2.8)
\put( 0.0,0.0){\includegraphics[width=4.cm,height=2.8cm]{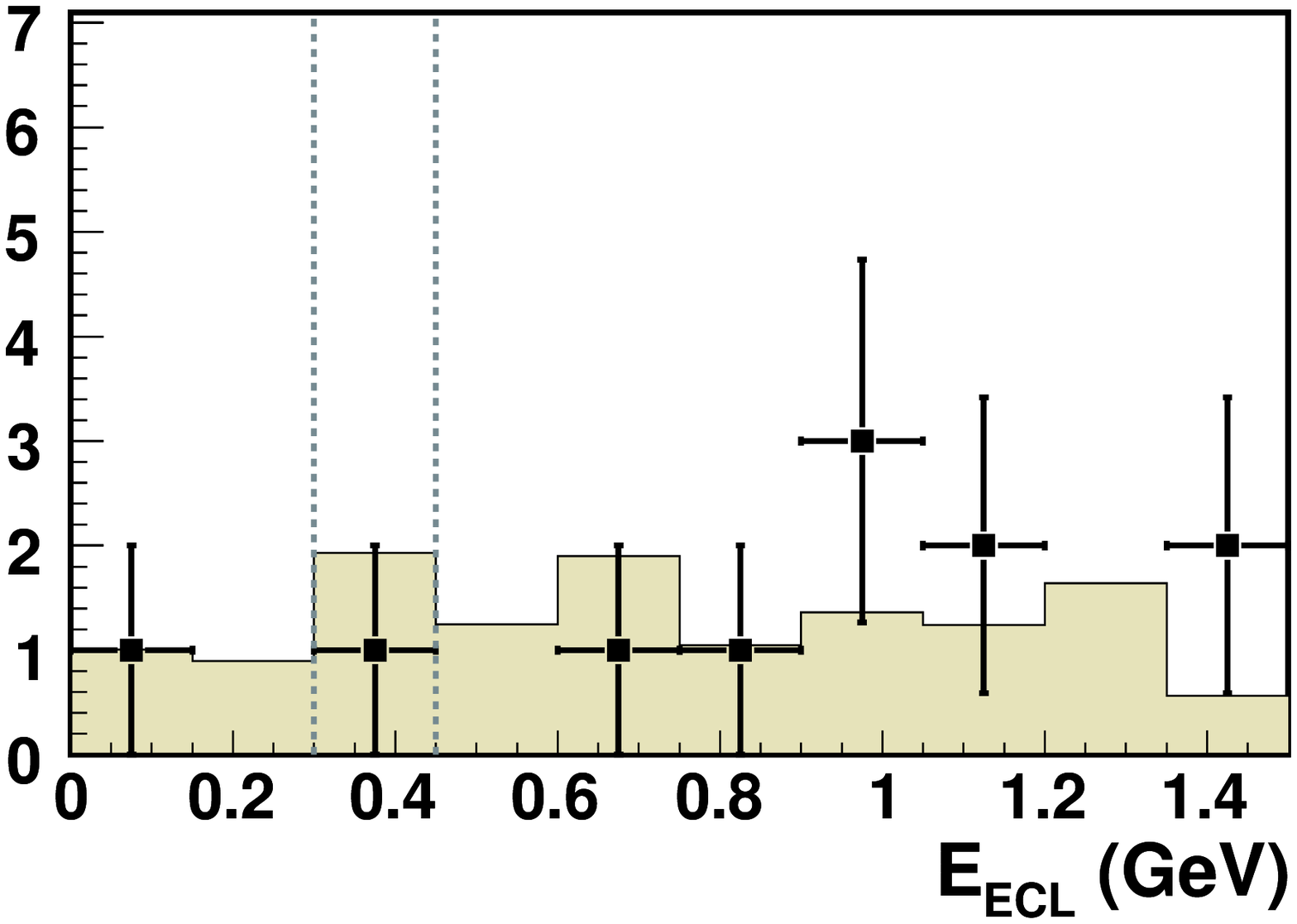}}
\put( 1.3,2.3){\footnotesize i) $B^0 \to \phi \nu \overline{\nu}$}
\end{picture}
\end{center}
\caption{The $E_{\mathrm{ECL}}$ distributions for $B \to h^{(*)}\nu\overline{\nu}$ decays.
The shaded histograms show the background distributions from MC
simulations and are normalized to sideband data. 
The open histograms show the SM expected signal distributions for
$B \to K^{(*)}\nu\overline{\nu}$ decays multiplied by a factor of 20 for the comparison.
The vertical dashed lines show the upper bound (left) 
of the signal box and the lower bound (right) of the sideband region.}
\label{fig:hnunubar}
\end{figure}

The data $E_{\mathrm{ECL}}$ distributions  are shown in Figure~\ref{fig:hnunubar}.
The distributions of background are estimated with 
MC simulations and are normalized by the number of events in the sideband region.
None of the signal modes has a significant signal. Including the effects of both statistical and
systematic uncertainties, an extension of the Feldman-Cousins 
method~\cite{Feldman:1997qc,Conrad:2002kn} is used to calculate the upper limits.
The observed number of events in the signal box and sideband region, expected background 
contributions in the signal box, reconstruction efficiencies, and the obtained upper limits 
at 90\% confidence level (CL) are shown in Table~\ref{tab:hnunubar}.
The reconstruction efficiencies are estimated with MC simulations using the $B \to h^{(*)}$ form factors from 
Ref.~\cite{ref:form_factors}. 
The $B^0 \to \phi\nu\overline{\nu}$ MC samples are generated with the $B \to K^{*}$ form factors.

\begin{table}[htpb]
\caption{A summary of the number of observed events in the signal box ($N_{\rm obs}$), 
expected background yields ($N_{b}$) in the signal box, reconstruction efficiencies including both $B_{\rm tag}$ and $B_{\rm sig}$ ($\epsilon$), 
and the upper limits (U.L.) at 90\% CL.}
\label{tab:hnunubar}
\begin{center}
\begin{tabular}{|l|r|r|r|r|}
\hline
Mode~ & $N_{\rm obs}$ & $N_{b}$~~ & $\epsilon (\times 10^{-5})$ & U.L.~~ \\ 
\hline
$K^{*0}\nu\overline{\nu}$ &  7 &   $4.2\pm1.4$ &  $5.1\pm0.3$ & ~$<3.4\times 10^{-4}$ \\
$K^{*+}\nu\overline{\nu}$ &  4 &   $5.6\pm1.8$ &  $5.8\pm0.7$ & ~$<1.4\times 10^{-4}$ \\
$~~~\to K_S^0\pi^+$       &  1 &   $2.3\pm1.2$ &  $2.8\pm0.3$ &  \\
$~~~\to K^+\pi^0$         &  3 &   $3.3\pm1.4$ &  $3.0\pm0.4$ &  \\
$K^{+}\nu\overline{\nu}$  & 10 &  $20.0\pm4.0$ & $26.7\pm2.9$ & ~$<1.4\times 10^{-5}$ \\
$K^0\nu\overline{\nu}$    &  2 &   $2.0\pm0.9$ &  $5.0\pm0.3$ & ~$<1.6\times 10^{-4}$ \\
$\pi^+\nu\overline{\nu}$  & 33 &  $25.9\pm3.9$ & $24.2\pm2.6$ & ~$<1.7\times 10^{-4}$ \\
$\pi^0\nu\overline{\nu}$  & 11 &   $3.8\pm1.3$ & $12.8\pm0.8$ & ~$<2.2\times 10^{-4}$ \\
$\rho^0\nu\overline{\nu}$ & 21 &  $11.5\pm2.3$ &  $8.4\pm0.5$ & ~$<4.4\times 10^{-4}$ \\
$\rho^+\nu\overline{\nu}$ & 15 &  $17.8\pm3.2$ &  $8.5\pm1.1$ & ~$<1.5\times 10^{-4}$ \\
$\phi\nu\overline{\nu}$   &  1 &   $1.9\pm0.9$ &  $9.6\pm1.4$ & ~$<5.8\times 10^{-5}$ \\
\hline
\end{tabular}
\end{center}
\end{table}

The possible disagreement in the $E_{\rm ECL}$ distributions
between data and MC is checked using wrong-flavor combinatorial events, and an uncertainty of 0.1--2.0 events is included. 
Background contributions from rare $B$ decays are examined using a large MC sample and 
the variation in the background yield (0.1--1.8 events) is included as a systematic uncertainty.
The uncertainties in $B_{\mathrm{tag}}$ reconstruction (2.0\% for $B^0$ and 9.9\% for $B^\pm$) 
are estimated by comparing the yields of data and MC from the $B_{\mathrm{tag}}$ candidates.
Systematic uncertainty arising from the track and $\pi^0$ rejection is studied using $B \to D^{(*)}\ell\nu$
decays, and an error of 2.7\% is assigned. 
The uncertainties in the efficiencies including detecting a $K_S^0$ (4.9\%) or $\pi^0$ (4.0\%),
$B \to h^{(*)}$ form factors (0.4--13\%), the number of $B\overline{B}$ events (1.3\%), 
tracking efficiency (1.0--2.2\%), particle identification (0.5--2.0\%), $h^{(*)}$ mass selection (0.8--2.3\%), and
the $\phi\to K^+K^-$ branching fraction (1.2\%).

We have performed a search for  
$B \to h^{(*)} \nu \overline{\nu}$ decays with a fully reconstructed $B$ tagging method on a
data sample of $535\times 10^{6}$ $B\bar{B}$ pairs. 
No significant signal is observed and we set upper limits on the branching fractions
at 90\% CL. 
The limits obtained for $B^0 \to K^{*0}\nu\overline{\nu}$ and $B^+ \to K^+\nu\overline{\nu}$ decays
are more stringent than the previous constraints.
The first searches for
$B^0 \to K^0 \nu\overline{\nu}$, $\pi^0\nu\overline{\nu}$, $\rho^0\nu\overline{\nu}$, $\phi\nu\overline{\nu}$, and
$B^+ \to K^{*+}\nu\overline{\nu}$, $\rho^+\nu\overline{\nu}$ are carried out.
The results still allow room for substantial non-SM contributions, thus
a higher luminosity $B$-factory experiment is required to probe the SM predictions for the branching fractions.

\begin{acknowledgments}
We thank the KEKB group for excellent operation of the
accelerator, the KEK cryogenics group for efficient solenoid
operations, and the KEK computer group and
the NII for valuable computing and Super-SINET network
support.  We acknowledge support from MEXT and JSPS (Japan);
ARC and DEST (Australia); NSFC and KIP of CAS (China); 
DST (India); MOEHRD, KOSEF, KRF and SBS Foundation (Korea); 
KBN (Poland); MES and RFAAE (Russia); ARRS (Slovenia); SNSF (Switzerland); 
NSC and MOE (Taiwan); and DOE (USA).
\end{acknowledgments}

\bigskip 

\end{document}